\documentclass[11pt]{article}
\usepackage{amsmath,amssymb,bm}
\usepackage{graphics}
\usepackage{graphicx}
\usepackage{amsfonts}
\usepackage{amssymb}
\usepackage{epstopdf}

\usepackage{mystyle}
\usepackage{cite,./mcite}
\usepackage{authblk}
\usepackage{lineno}
\switchlinenumbers

                \def\lsim{\mathrel{\rlap{\lower4pt\hbox{\hskip1pt$\sim$}}
    \raise1pt\hbox{$<$}}}                \def\gsim{\mathrel{\rlap{\lower4pt\hbox{\hskip1pt$\sim$}}
    \raise1pt\hbox{$>$}}}

\title{Towards Diffraction in Herwig}
\author[1]{Stefan Gieseke}
\author[1]{Frash\"{e}r Loshaj}
\author[2]{Miroslav Myska}
{\tiny
\affil[1]{Institute for Theoretical Physics, Karlsruhe Institute of Technology}
\affil[2]{Czech Technical University in Prague, FNSPE, Brehova 7, Prague}
}

\begin{document}

\maketitle 
\begin{abstract}
We propose changes to the colour reconnection model in the Monte Carlo event generator Herwig in order to remove the quasi diffractive events from the soft multiple parton interactions. We then implement explicitly soft diffraction and show some preliminary results. 
\end{abstract}

The multiple parton interactions (MPI) models in Monte Carlo generators are crucial in explaining many observables in hadron collisions. For a recent review of Monte Carlo generators see \cite{Buckley:2011ms}. One can view the incoming hadrons as bunches of incoming partons, where many scatterings between them can take place. 
One of the motivations for considering the MPI is the fact that the inclusive jet cross section, calculated in the usual perturbative QCD formalism, exceeds the total cross section at moderate values of the center of mass energy (see for example \cite{Bahr:2009ek}). A resolution to this problem is to use MPI, where one considers the luminosity of collided particles \cite{Borozan:2002fk,Bahr:2008wk} to factorize in an impact parameter dependent factor and the usual PDFs. One can parametrize the average multiplicity of multiple hard scatterings as a function of the cross section and an impact parameter dependent function - the overlap function. One can then use the eikonal model to parametrize the amplitude of multiple scattering as is done in Herwig \cite{Bahr:2008dy}.

\begin{figure}
        	\centering
        		\includegraphics[width=.8\linewidth]{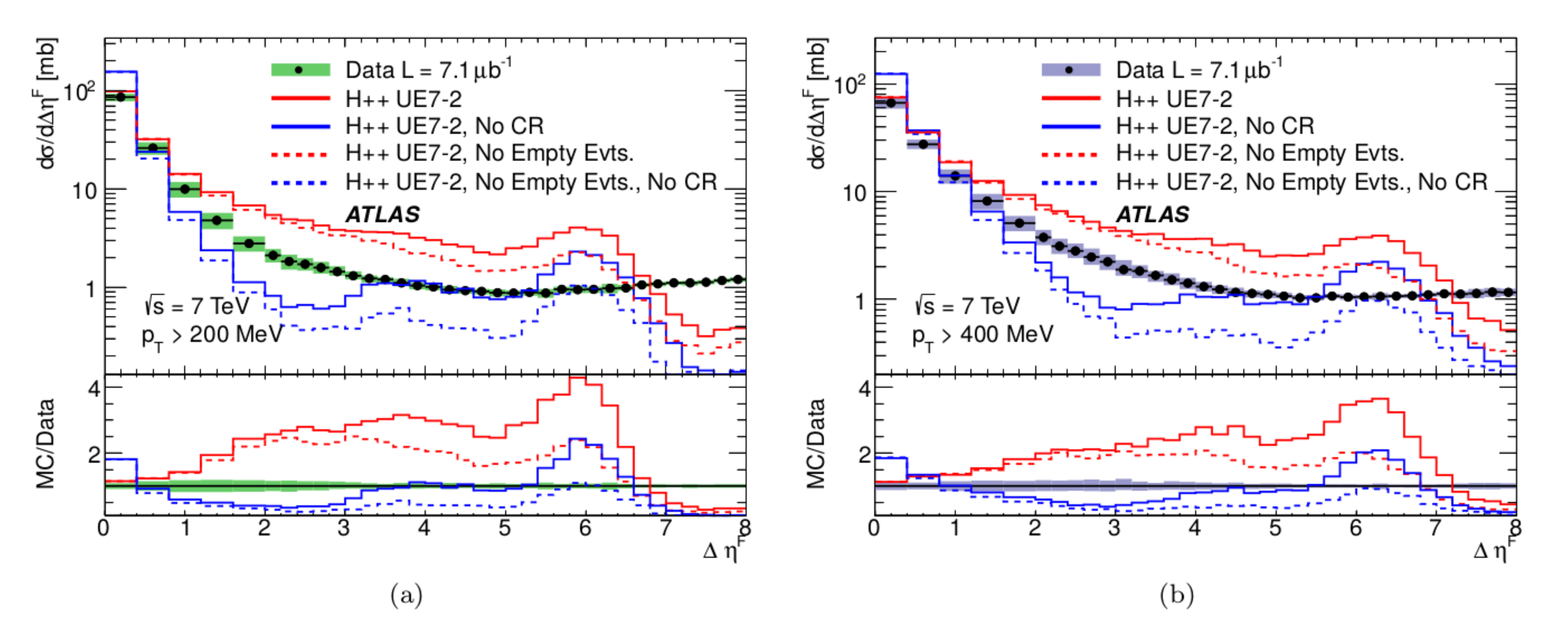}
        		\caption{Inelastic differential cross section versus forward pseudorapidity gap $\Delta\eta^F$.}
        		\label{fig:atlas}
        \end{figure}

Soft MPI in Herwig is simulated using the eikonal model as well. The kinematics of the outgoing soft partons is implemented by generic two-to-two processes with colour connections between them and the rest of the process explicitly specified. The purpose of this model was to describe minimum bias data at central pseudorapidity and small pseudorapidity gaps. Although the model was not expected to describe events with large pseudorapidity gaps, it was surprising to see that it gives an enhancement, rather than a depletion of those events. This can be seen in a recent measurement by ATLAS \cite{Aad:2012pw}, shown in Fig. \ref{fig:atlas}, where the differential cross section versus the forward pseudorapidity is shown. The forward pseudorapidity gap is defined as the larger of two pseudorapidities from the last particle to the edge of the detector. 

The enhancement seems to be larger when the colour reconnection model is switched on. The colour reconnection model in Herwig creates different cluster configurations from the original colour topology in order to minimize the sum of the cluster masses \cite{Gieseke:2012ft}. This procedure seems to produce large pseudorapidity gaps, partly due to the splitting of large clusters into smaller ones, which end up being too forward in pseudorapidity. 

In what follows we will address this issue. It is well known that the large pseudorapidity gaps should come from diffractive events. The differential cross section in $\Delta \eta^F$, using Regge phenomenology, should behave like $\sim e^{\Delta\eta^F(\alpha_\mathbb{P}(0)-1)}$ (see \cite{Barone:2002cv} for a review), where $\alpha_\mathbb{P}(0)$ is the pomeron intercept and its value is close to 1. This explains the almost constant value at large $\Delta \eta^F$. The differential cross section for nondiffractive events (at small $\Delta\eta^F$) behaves like $\sim e^{-\Delta\eta^F}$. 

\begin{figure}
			\centering
         	\includegraphics[width=.4\linewidth]{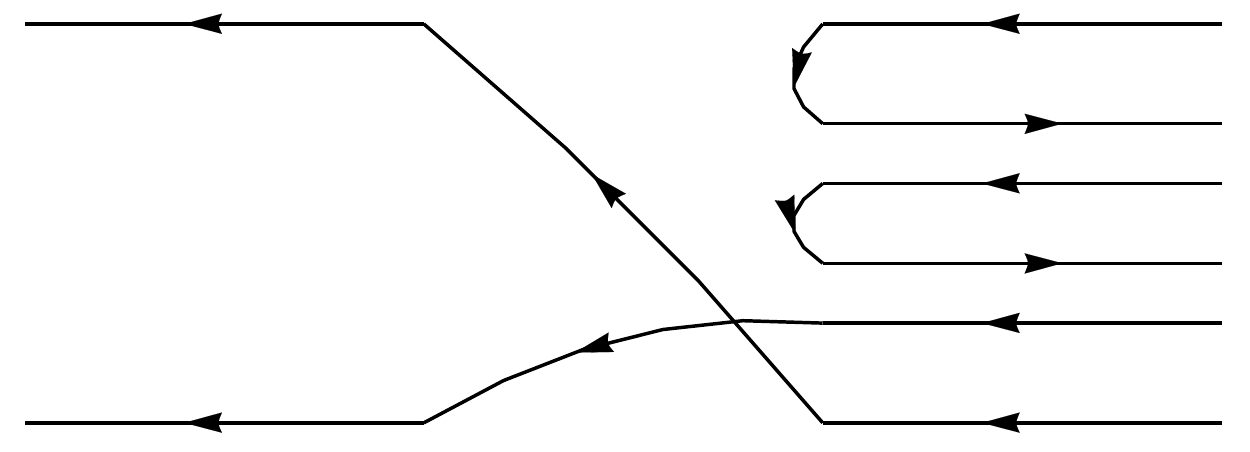}
         	\includegraphics[width=.4\linewidth]{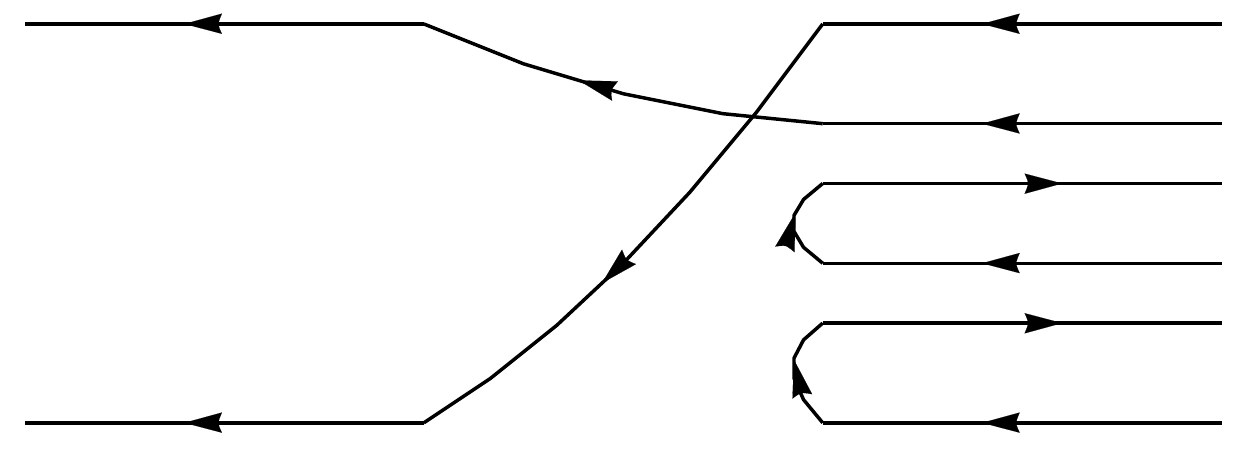}
         	\caption{Colour connection topologies with small pseudorapidity gaps.}
         	\label{fig:topol}
		\end{figure}

In order to get the proper behaviour of the cross section we have to consider separately the non-diffractive and diffractive parts. To get the correct non-diffractive cross section, we have to modify the existing colour connections between the soft partons and proton remnants in the soft MPI model. An example of a topology giving the required suppression of large gaps is given in Fig. \ref{fig:topol}. 
		
\begin{figure}
\centering
\includegraphics[width=.4\linewidth]{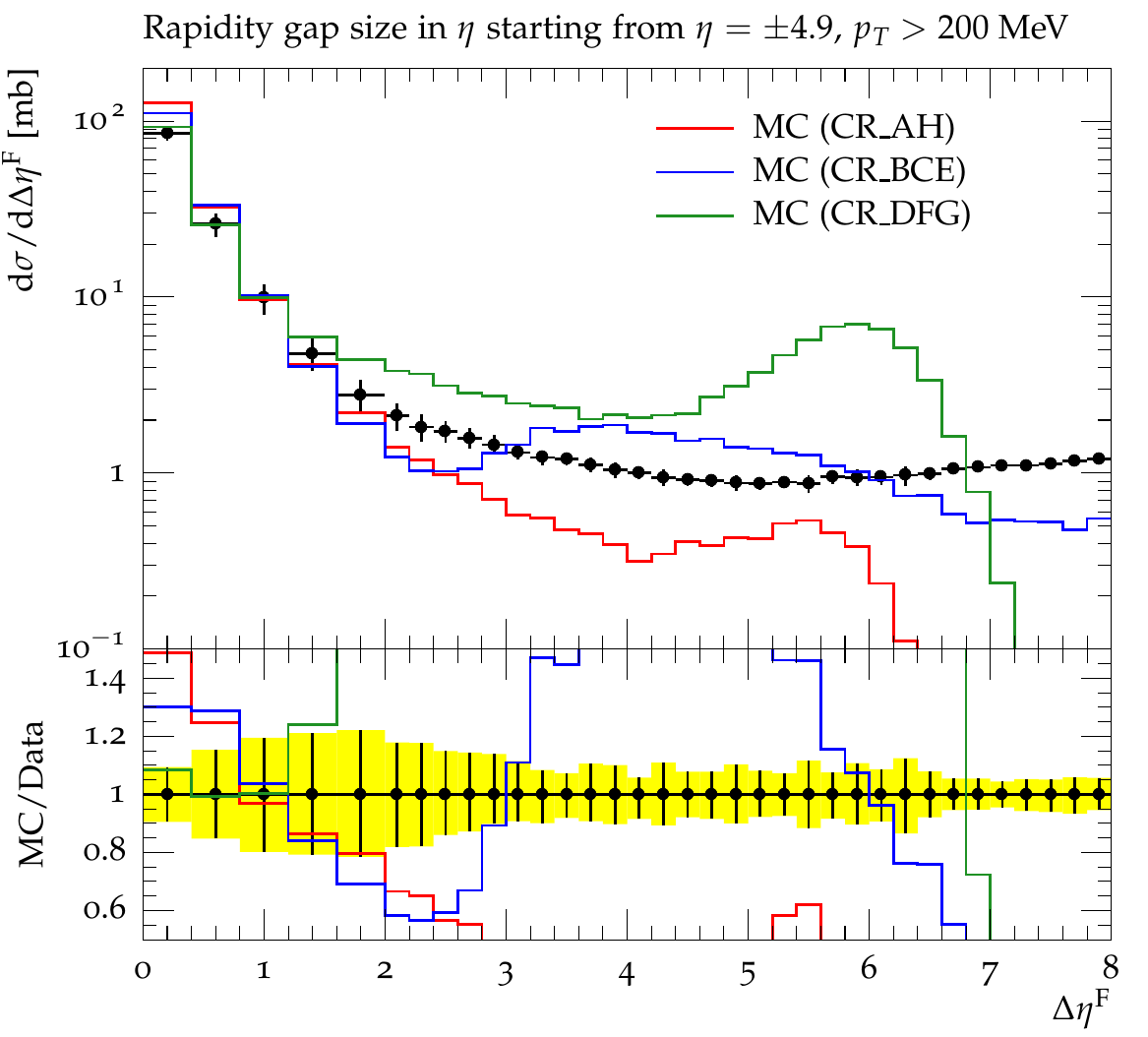}
\caption{Differential cross section in forward pseudorapidity gap $\Delta\eta^F$ in Herwig using different topologies of colour connections.}
\label{fig:minbias}
\end{figure}

Arrows pointing to the (left) right, represent an outgoing (anti) colour line. The top and bottom lines represent proton remnants which are connected to either soft partons or the other remnant. 
Contributions from different classes of colour topologies denoted by letters A,H, etc, are shown in Fig. \ref{fig:minbias}. We notice that none of the topologies gives the required exponential fall off. This behaviour may be obtained by changes in the colour reconnection model in restricting reconnection of forward clusters, but also more work is needed to understand better the colour connections themselves. This concludes our discussion of the non-diffractive cross section.

\begin{figure}
		\centering
        \includegraphics[width=.4\linewidth]{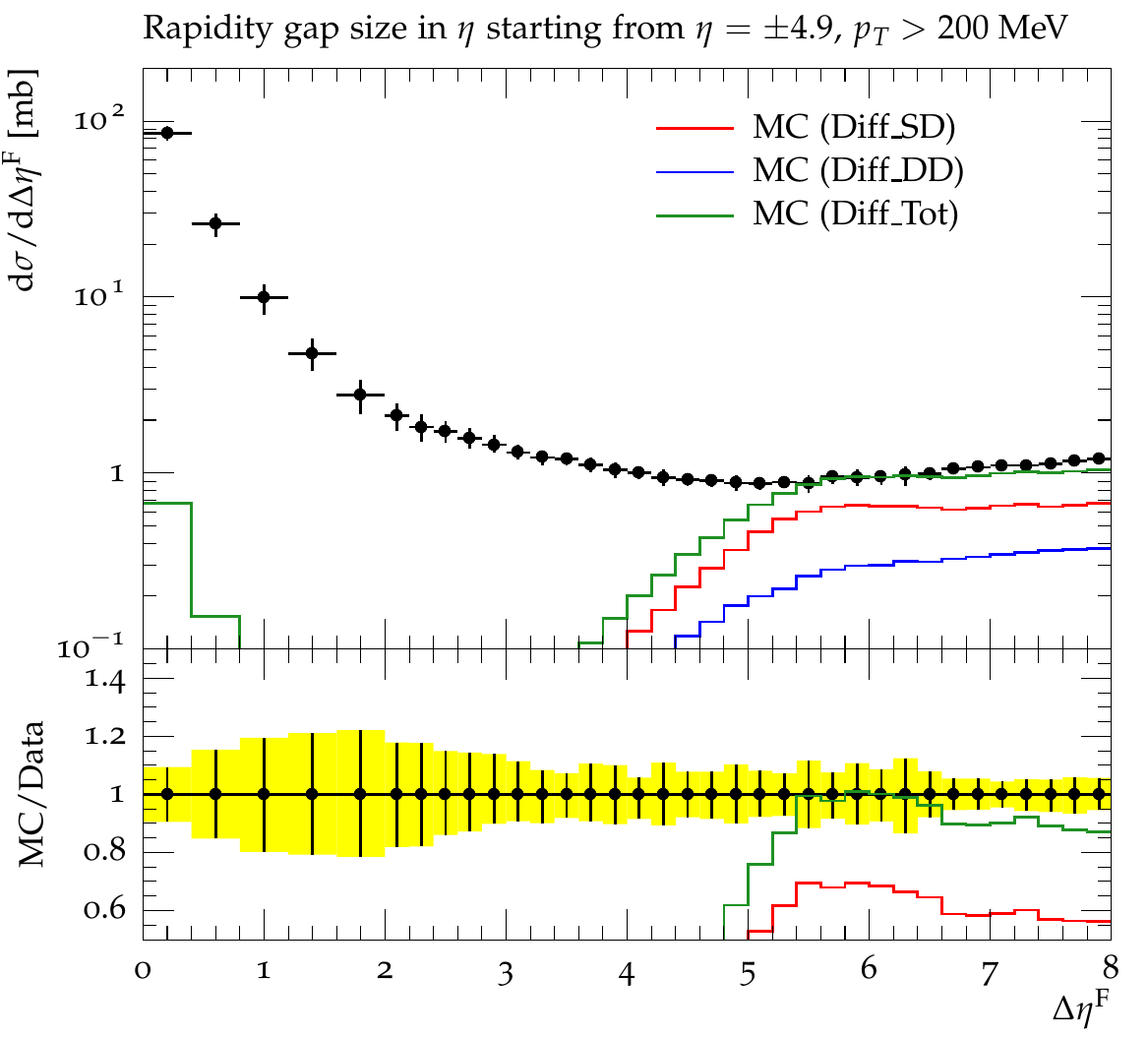}
        \includegraphics[width=.4\linewidth]{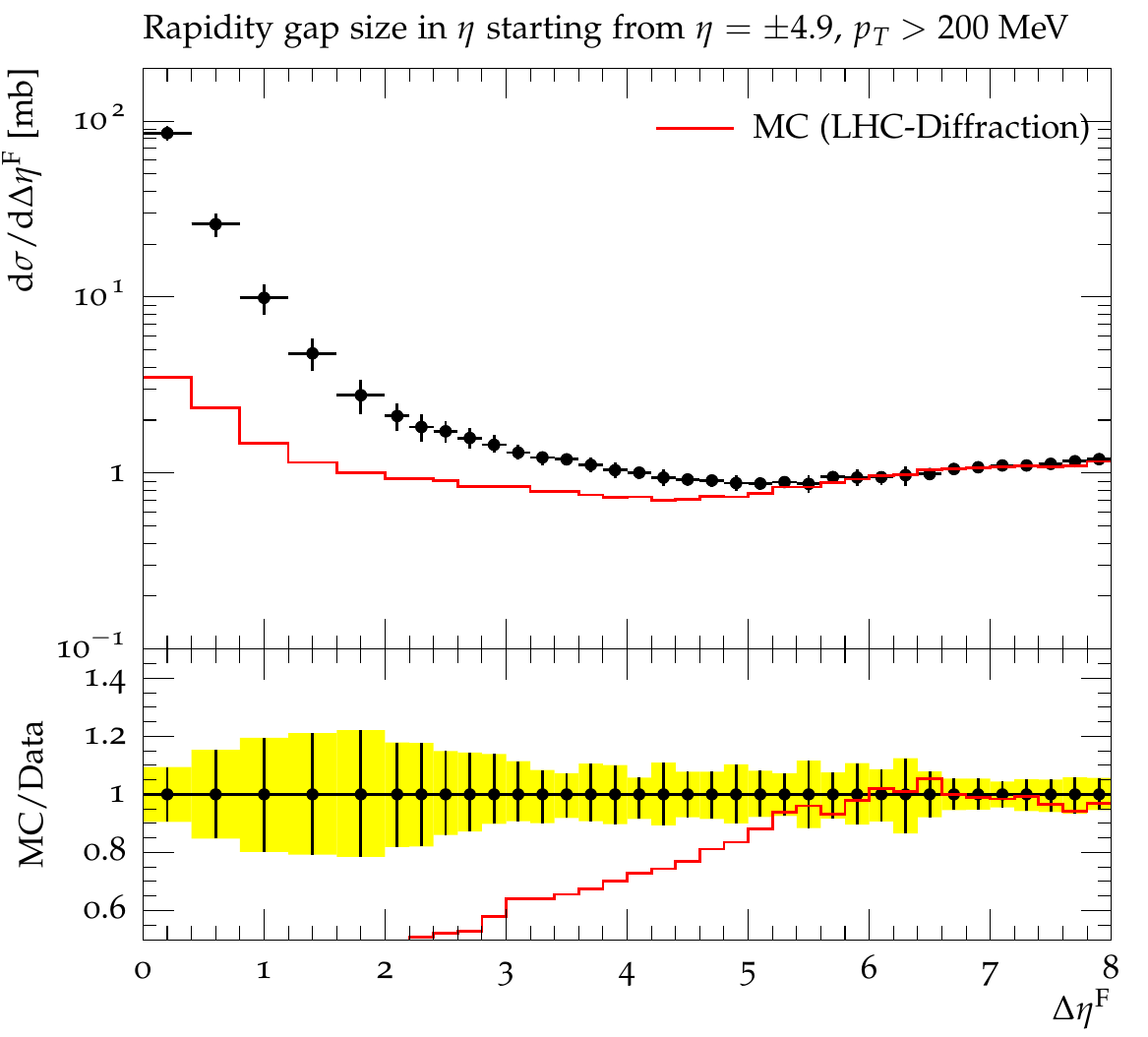}
        \caption{Single and double diffractive contribution to differential cross section in forward pseudorapidity gap. Left: diffraction with isotropic decay of dissociated proton. Single and double diffraction shown separately. Right: sum of single and double diffractive events in the case when quark and diquark are collinear with outgoing dissociated proton. }
        \label{fig:diff}	
	\end{figure}

We next move to the generation of diffractive events. Diffraction is implemented using the results from Regge theory and the cross sections are given in the familiar form, for single diffraction 
\begin{align}
\frac{d^2\sigma^{SD}}{dM^2dt}=\frac{1}{16\pi^2 s}|g_\mathbb{P}(t)|^2 g_\mathbb{P}(0) g_{\mathbb{PPP}}(0) \left(\frac{s}{M^2}\right)^{2\alpha_\mathbb{P}(t)-1} \left(M^2 \right)^{\alpha_\mathbb{P}(0)-1},
\end{align}
and double diffraction
\begin{align}
\frac{d^3\sigma^{DD}}{dM_1^2dM_2^2dt}=\frac{1}{16\pi^3 s} g_\mathbb{P}^2(0) g_{\mathbb{PPP}}^2(0) \left(\frac{s}{M_1^2M_2^2}\right)^{2\alpha_\mathbb{P}(t)-1}  \left(M_1^2\right)^{\alpha_\mathbb{P}(0)-1}\left(M_2^2\right)^{\alpha_\mathbb{P}(0)-1},
\end{align} 
where $\alpha_\mathbb{P}(t)=\alpha_\mathbb{P}(0)+\alpha't$ with $\alpha_\mathbb{P}(0)$ and $\alpha'$ being the pomeron intercept and slope respectively, and $g_\mathbb{P}=g_{p\mathbb{P}}$ is the proton pomeron coupling.
The events are generated using a matrix element which implements the two-to-two body kinematics from the cross sections above. Namely, we implement processes $pp\rightarrow p^*p$, $pp\rightarrow pp^*$ and $pp\rightarrow p^*p^*$, where $p^*$ is the dissociated proton. The dissociation is simulated by a cluster made up of a quark and a diquark. We first consider an isotropic decay of the dissociated proton into quark and diquark. The preliminary result is shown in Fig. \ref{fig:diff}, left. In this case we notice a quick fall off of the cross section for small $\Delta \eta^F$, which should be constant for a significant range. We also consider the case when the quark is collinear with the outgoing dissociated proton. In this case the hadronization effects populate the region of small forward pseudorapidity gap as shown in Fig. \ref{fig:diff}, right. The cluster hadronization seems to give better results in this case, but not exactly as expected, since the constant behaviour of the cross section doesn't extend to low values of $\Delta \eta^F$ and hadronization effects start at very high $\Delta\eta^F$.  \ref{fig:diff}. 
An improvement could come from considering the dissociation into quark, diquark and gluon, since this scenario allows for more isotropic distribution of final state particles and it could shift hadronization effects at very low $\Delta\eta^F$ as expected.		

In conclusion, we have shown that large pseudorapidity gaps appearing in minimum bias results in Herwig can be tamed by introducing new colour connection topologies for soft interactions and modifying the colour reconnection model. Events with final state particles in the forward pseudorapidity region should be described properly using diffraction. We have shown how one can implement diffraction in Herwig and some preliminary results.
\section*{Acknowledgements}
The authors are grateful to A. Siodmok and M. Seymour for useful discussions.  This work was supported in part by the European Union as part of the FP7 Marie Curie Initial Training Network MCnetITN (PITN-GA-2012-315877). 

\bibliographystyle{MPI2015}
\bibliography{references}

\end{document}